\documentclass[a4paper,11pt]{article}
\pdfoutput=1 
\usepackage{jinstpub} 
\usepackage{lineno}

\usepackage{siunitx}
\usepackage{subcaption}
\usepackage{graphicx}
\usepackage{textcomp}

\title{\boldmath{Quality assessment of Cadmium Telluride as a detector material for multispectral medical imaging}\\ \vspace{1em}\small\textit
{
This is the Accepted Manuscript version of an article accepted for publication in \textit{Journal of Instrumentation}. IOP Publishing Ltd is not responsible for any errors or omissions in this version of the manuscript 
or any version derived from it. The Version of Record is available online at \url{https://doi.org/10.1088/1748-0221/17/01/C01070}.}
 }

\author[a,1]{S.~Kirschenmann,\note{Corresponding author.}}
\author[a,b]{M.~Bezak,}
\author[a]{S.~Bharthuar,}
\author[a,g]{E.~Br{\"{u}}cken,}
\author[a,b]{M.~Golovleva,}
\author[a,f]{A.~G{\"{a}}dda,}
\author[a]{M.~Kalliokoski,}
\author[a,h]{A.~Karadzhinova-Ferrer,}
\author[a]{P.~Koponen,}
\author[a,b]{N.~Kramarenko,}
\author[a,b]{P.~Luukka,}
\author[a,d]{J.~Ott,}
\author[c, a]{J.~Tikkanen}
\author[a]{R.~Turpeinen}
\author[e, a]{and A.~Winkler}

\affiliation[a]{Helsinki Institute of Physics, Gustaf H{\"{a}}llstr{\"{o}}min katu 2, FI-00014 University of Helsinki, Finland}
\affiliation[b]{Lappeenranta University of Technology, Skinnarilankatu 34, FI-53850 Lappeenranta, Finland}
\affiliation[c]{Radiation and Nuclear Safety Authority (STUK), Laippatie 4, FI-00880 Helsinki, Finland}
\affiliation[d]{Santa Cruz Institute for Particle Physics, University of California Santa Cruz,
Santa Cruz, CA 95064, USA}
\affiliation[e]{Detection Technology Plc, A Grid, Otakaari 5a, 02150 Espoo, Finland}
\affiliation[f]{Okmetic Oy, Piitie 2, FI-01510 Vantaa, Finland}
\affiliation[g]{Xiangtan University, Hunan, China}
\affiliation[h]{Ludong University, Yantai, China}

\emailAdd{stefanie.kirschenmann@helsinki.fi}

\abstract{
Cadmium Telluride (CdTe) is a high-Z material with excellent photon radiation absorption properties, making it a promising material to include in radiation detection technologies. However, the brittleness of CdTe crystals as well as their varying concentration of defects necessitate a thorough quality assessment before the complex detector processing procedure.
We present our quality assessment of CdTe as a detector material for multispectral medical imaging, a research which is conducted as part of the Consortium Project Multispectral Photon-counting for Medical Imaging and Beam characterization (MPMIB). The aim of the project is to develop novel CdTe detectors and obtain spectrum-per-pixel information that make the distinction between different radiation types and tissues possible. \\
To evaluate the defect density inside the crystals -- which can deteriorate the detector performance -- we employ infrared microscopy (IRM). Posterior data analysis allows us to visualise the defect distributions as 3D defect maps. Additionally, we investigate front and backside differences of the material with current-voltage (IV) measurements to determine the preferred surface for the pixelisation of the crystal, and perform test measurements with the prototypes to provide feedback for further processing. We present the different parts of our quality assessment chain and will close with first experimental results obtained with one of our prototype PC detectors in a small tomographic setup.
}


\proceeding{22$^{\text{nd}}$ International Workshop on Radiation Imaging Detectors\\
  27 June 2021 to 1 July 2021\\
  Online}

\begin{document}
\maketitle
\flushbottom

\section{Introduction}
\label{sec:intro}
The presented research is part of the project \textbf{M}ultispectral \textbf{P}hoton-counting for \textbf{M}edical \textbf{I}maging and \textbf{B}eam characterization (MPMIB), funded via the Academy of Finland RADDESS programme~\cite{RADDESS}. The MPMIB project focuses on the development of a next-generation radiation detection system operating in a photon-counting (PC) mode~\cite{Brucken2020}. The goal is the extraction of spectrum-per-pixel data, which will lead to higher efficiency and image quality, as well as the possibility to identify different materials and tissue types. Our approach is to fabricate direct-conversion semiconductor detectors hybridized with PC capable read-out chips (ROC). Currently, we focus on solutions with Cadmium Telluride (CdTe), which is a high-Z material with excellent photon radiation absorption properties. \\
The processing of the CdTe material into detectors is performed at Micronova facilities in Espoo, Finland, using bare $10\times10\times1~\mathrm{mm}^3$ CdTe crystals obtained from Acrorad Ltd. Our detector processing includes i.a. employing atomic layer deposition (ALD) for depositing alumina ($\text{Al}_2\text{O}_3$) as thin film insulation material on the crystal, as well as sputtering titanium tungsten (TiW) as metal contacts. The pixelised crystals are flip-chip-bonded to the matching PSI46digV2.1-r ROC~\cite{Meier:2011zz, Hits:2015jsa}\footnote{Available to us due to our participation in the CMS experiment at CERN.}. The ROC has a layout of 52 columns and 80 rows, accounting for 4160 pixel connections, with a pixel size of 100$\times$150~$\mu$m$^2$. The processing procedure is described in more detail in~\cite{Brucken2020, Gadda2017, Gadda2019}. 
However, the processing of Cadmium Telluride is more complicated than for standard materials such as silicon: it is a brittle material and we face problems such as cracking damages during processing, delamination of the insulation material, differences between the front and back sides of the crystals, and high leakage current of the manufactured detector. For these reasons, a quality assessment prior to the complex detector manufacturing process is necessary. \\
The methods employed for this assessment will be described in the following. They include Infrared Microscopy (IRM) for the visualisation of the defect density inherent in the CdTe crystals~\cite{Kirschenmann2021}, Current-Voltage (IV) measurements to investigate the electrical properties of the CdTe crystal and prototype testing.
We will discuss IRM results showing defect distributions of CdTe crystals and the possible implications on the performance of the processed detector, showcase our findings concerning the side-differences investigated with IV, as well as present first experimental data obtained with one of our prototype PC detectors in a small tomographic setup.

\section{Quality assessment of CdTe crystals with Infrared Microscopy}
\begin{figure}[htb]
  \centering
  \begin{subfigure}[b]{0.3\linewidth}
    \centering
     \includegraphics[width=\linewidth]{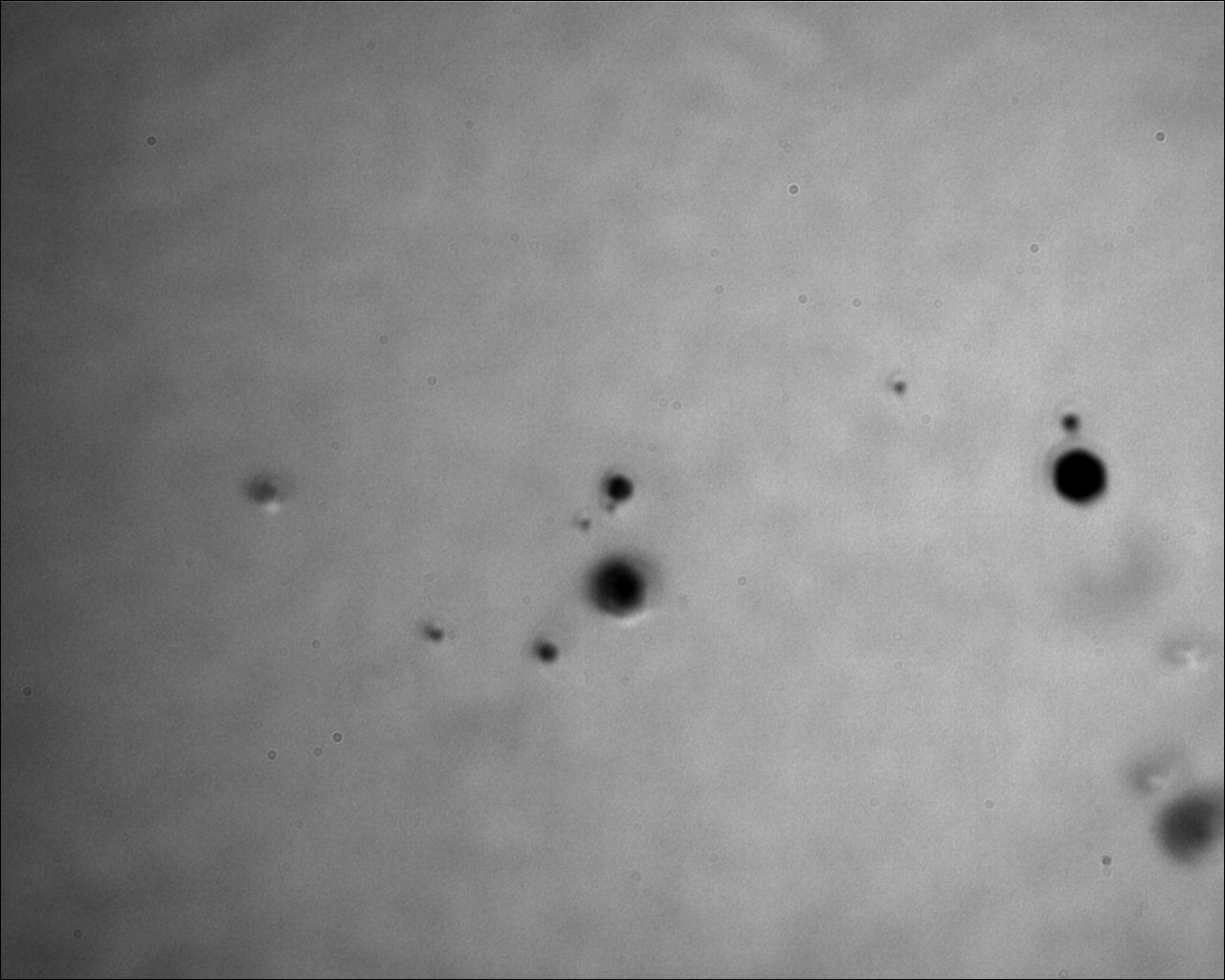}
    \caption{}
  \end{subfigure}
  \hspace{1em}
  \begin{subfigure}[b]{0.3\linewidth}
    \centering
    \includegraphics[width=\linewidth]{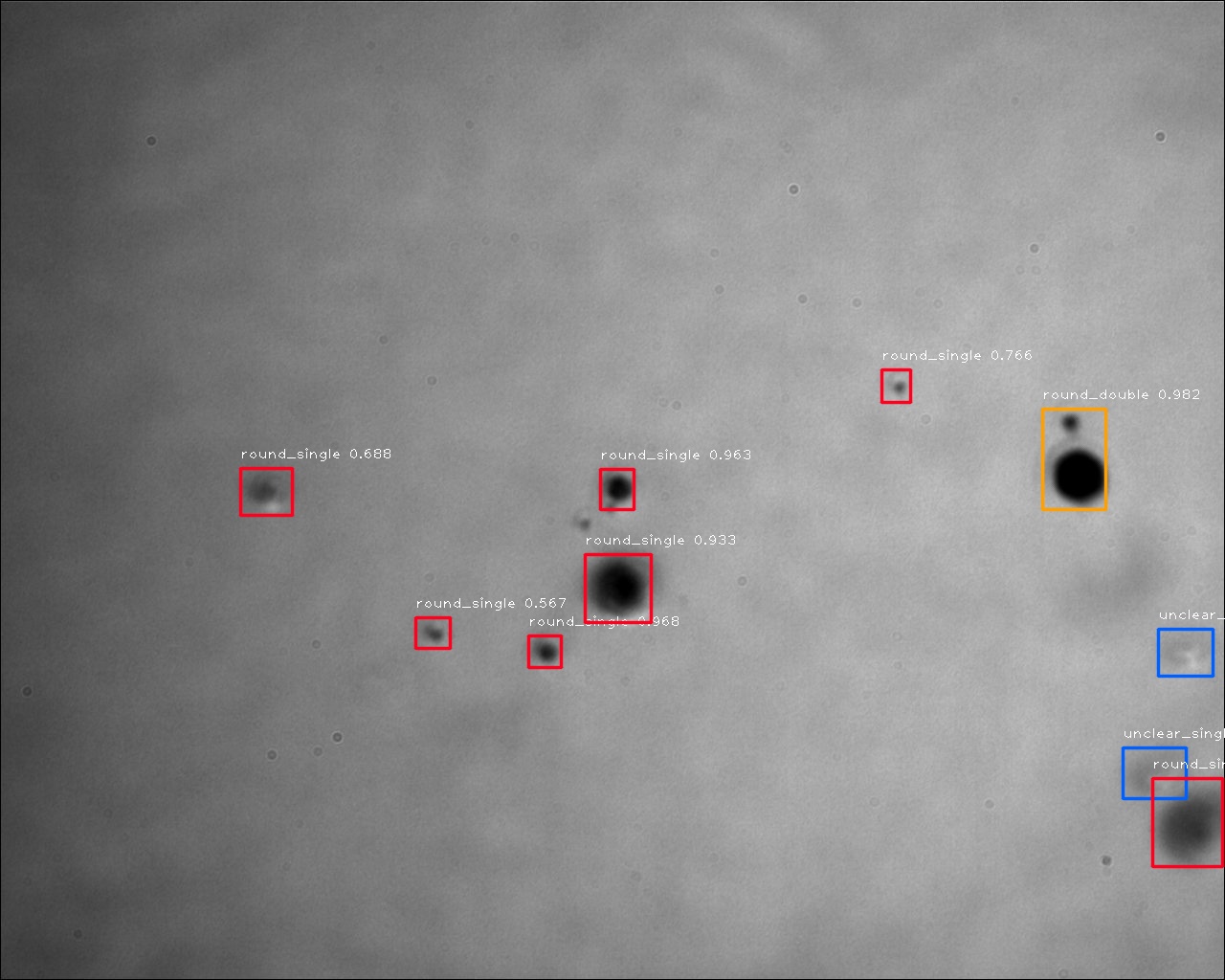}
    \caption{}
  \end{subfigure}
  \vspace*{-0.3cm}
  \caption{Example of IRM images evaluated by the employed neural network (FOV 200\,$\mu$m x 160\,$\mu$m),  a) showing the bare IRM image and b) the neural network classification.}
  \label{IRM}
\end{figure}
CdTe crystals can include larger amounts of defects, such as Tellurium (Te) inclusions and grain boundaries. These defects can lead to a deterioration in the processed detector~\cite{Bolotnikov2007, Bolotnikov2007b, Bolotnikov2008}.
As the CdTe compound is transparent for near-infrared light, but Te is opaque for light in this region, we can use infrared microscopy (IRM) to make Te inclusions (from now on: "defects") inside the material visible. We described the usage of a 3D IRM setup in combination with neural network detection and subsequent data analysis in~\cite{Kirschenmann2021}.
Fig.~\ref{IRM}a shows an example of an IRM image taken with the setup, with a Field of View (FOV) of $200\times160\,\mu m$ for a single image. Fig.~\ref{IRM}b shows the same image with annotations from the neural network, marking the detected objects. As in the current IRM procedure images are taken at adjacent positions along layers of different depths inside the CdTe crystal, we can obtain information on complete layers. Further data exploration as described in~\cite{Kirschenmann2021} makes it possible to visualise and study the defect density in the crystal.

\paragraph{IRM results}
\begin{figure}[htb]
    \centering
    \includegraphics[width=0.9\linewidth]{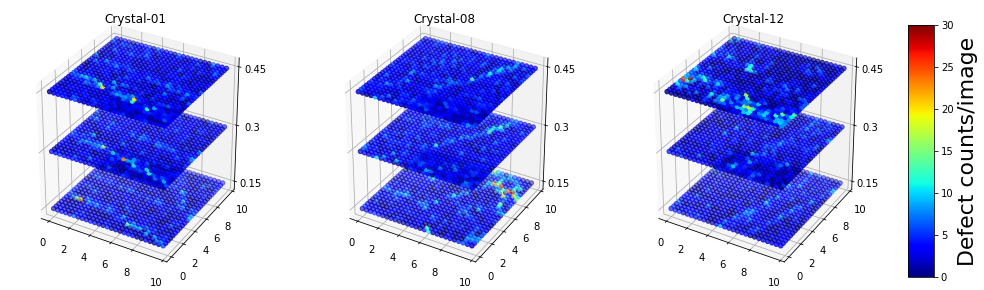}
  \vspace*{-0.3cm}
  \caption{3D defect maps of three different CdTe crystals (similarly presented in \cite{Kirschenmann2021}), axes in mm.}
  \label{IRMresults}
\end{figure}
The set of IRM images from a CdTe crystal is fed to a neural network for automatized defect detection. Further data analysis makes it possible to visualise the defect distribution inside the crystals in the form of 3D defect maps. Fig.~\ref{IRMresults} shows three defect maps, with each subplot standing for an individual CdTe crystal. Some local areas of higher defect density are clearly visible. These maps make it possible to compare crystals to each other with respect to their local defect densities. For further explanations on how these defect maps were produced and analysed, we refer to~\cite{Kirschenmann2021}. The defect density information can be used for a quality-sorting of the crystals and can be of interest, when comparing with the processed detector's performance. 


\section{Current-Voltage measurements}
The current-voltage (IV) measurements were performed at a probe station with a movable chuck and probe needles for connecting to the device under test. The probe station is connected to a Keithley 2410-C SourceMeter (supply of bias voltage), a Keithley 6487 PicoAmmeter (for registering the leakage current) and a measurement computer.
As high leakage current is detrimental for detector performance and operation, it is important to monitor the current.
In the following, we use the same naming for the crystal sides (\textit{A} and \textit{B}), as set by the manufacturer: the A side of the crystal is the side attached to an adhesive tape in the original sample box.
For the measurement of the metallized samples, we place the sample with the A side on the chuck of the probe station and place a probe needle on the top side (B), using micromanipulators. In case of the non metallized bare crystals, the contact is facilitated by attaching the crystal to an aluminum plane on top of a PCB with a probe head (as well aluminum). The leakage current is then read out via the attached probe needle on the B side, while applying a bias-voltage through the chuck to the back (A side) of the sample. 
\paragraph{IV results}
Measuring the IV curves of several CdTe test detectors and bare crystals, we followed a recurring phenomenon that the amount of the leakage current is dependent on the side from which the detector is biased. As we want to minimize the expected leakage current for the processed detector, it is important to know if we have a preferred side for the pixelisation.
\begin{figure}[htb]
  \centering
  \begin{subfigure}[b]{0.48\linewidth}
    \centering
     \includegraphics[width=\linewidth]{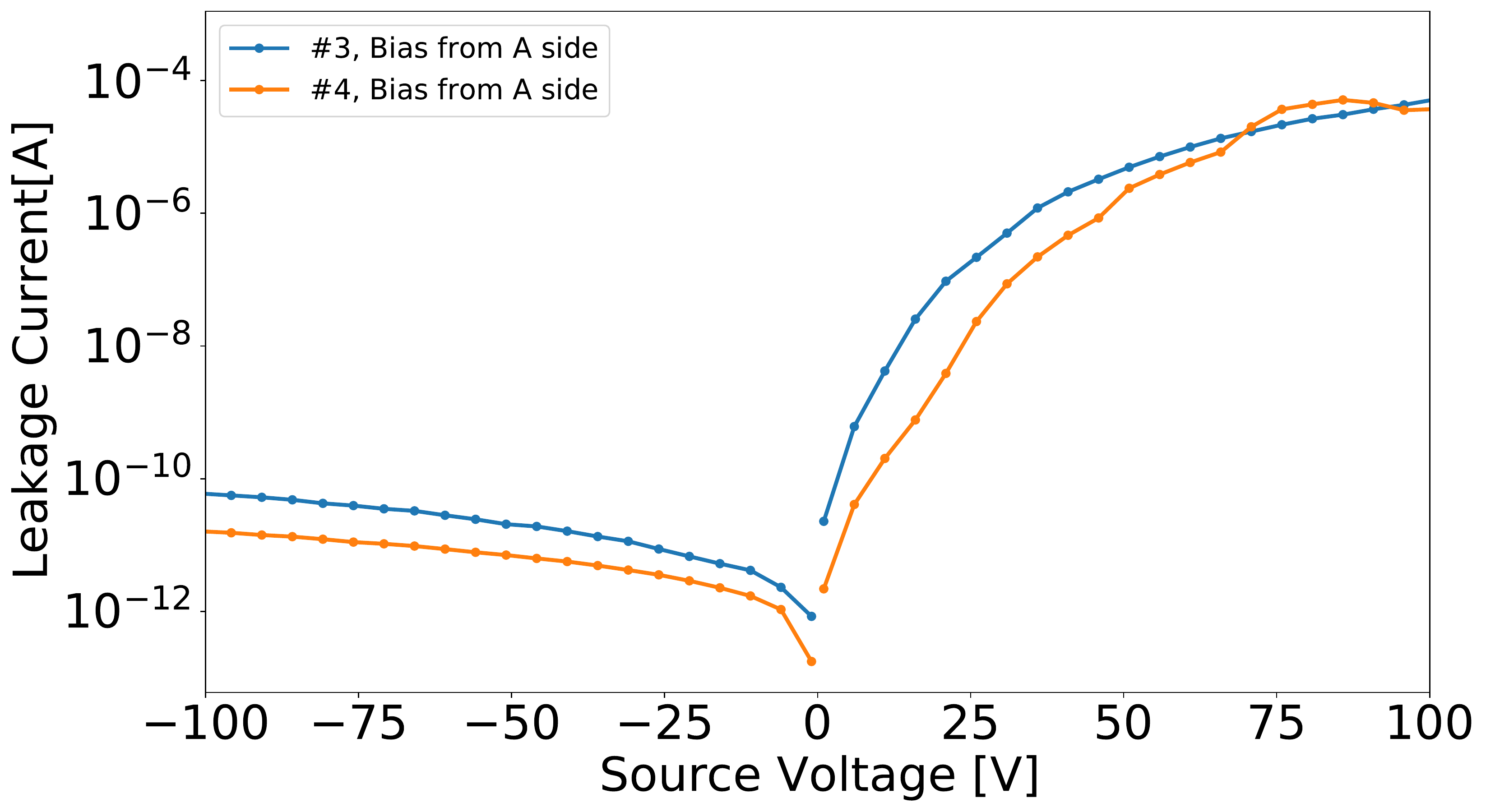}
    \caption{}
  \end{subfigure}
  \begin{subfigure}[b]{0.48\linewidth}
    \centering
    \includegraphics[width=\linewidth]{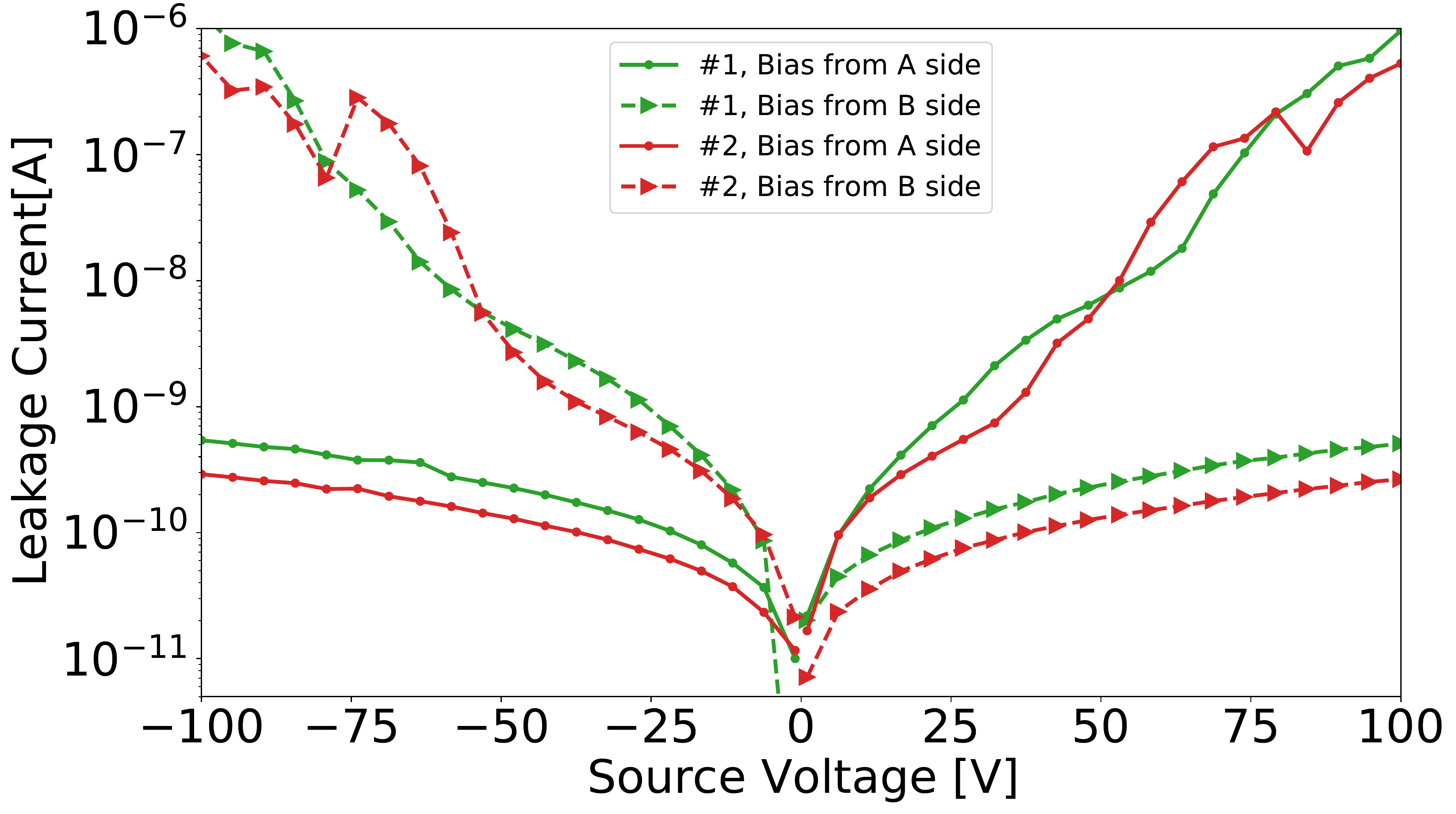}
    \caption{}
  \end{subfigure}
  \vspace*{-0.3cm}
  \caption{Asymmetric current-voltage behavior of CdTe crystals. (a) IV curves of two pixelised detector samples, (b) IV curves of two unprocessed CdTe crystals.}
  \label{IV-differences}
\end{figure}

In Fig.~\ref{IV-differences} we show exemplarily the IV curves for two bare CdTe crystals without any metallisation (1, 2) and two pixelised detectors (3, 4). Both for the non metallized CdTe crystals as well as the processed test detectors, a clear Schottky-type asymmetric current-voltage behaviour is visible, depending on whether we apply positive or negative bias. Furthermore, Fig.~\ref{IV-differences}b also shows the measurements with inversed sides, reading the current from the A side, while applying the bias voltage from the B side. The same Schottky-type behaviour can be noticed, in this case mirrored. As we want to collect electrons to the pixelised side (front), we need to apply a negative bias voltage from the backside. When biasing from side A, a clear preference for negative voltage can be observed both for the bare as well as the processed CdTe: We encounter leakage current differences of 3-6 orders of magnitude at a bias voltage of -100\,V and 100\,V, respectively. For collecting electrons, the B side is therefore the preferred side for pixelisation.  Observable differences in leakage current ranges between Fig.~\ref{IV-differences}a and b can be due to differences in the probing methods for current measurements as well as due to variations in the sample materials. \\
The reasons for the side differences is the subject of further investigation. For the time being, an IV-measurement step in the quality assessment and processing routine is retained.

\section{Spectral measurements}
\begin{figure}[htb]
  \centering
  \begin{subfigure}[b]{0.35\linewidth}
    \centering
     \includegraphics[width=\linewidth]{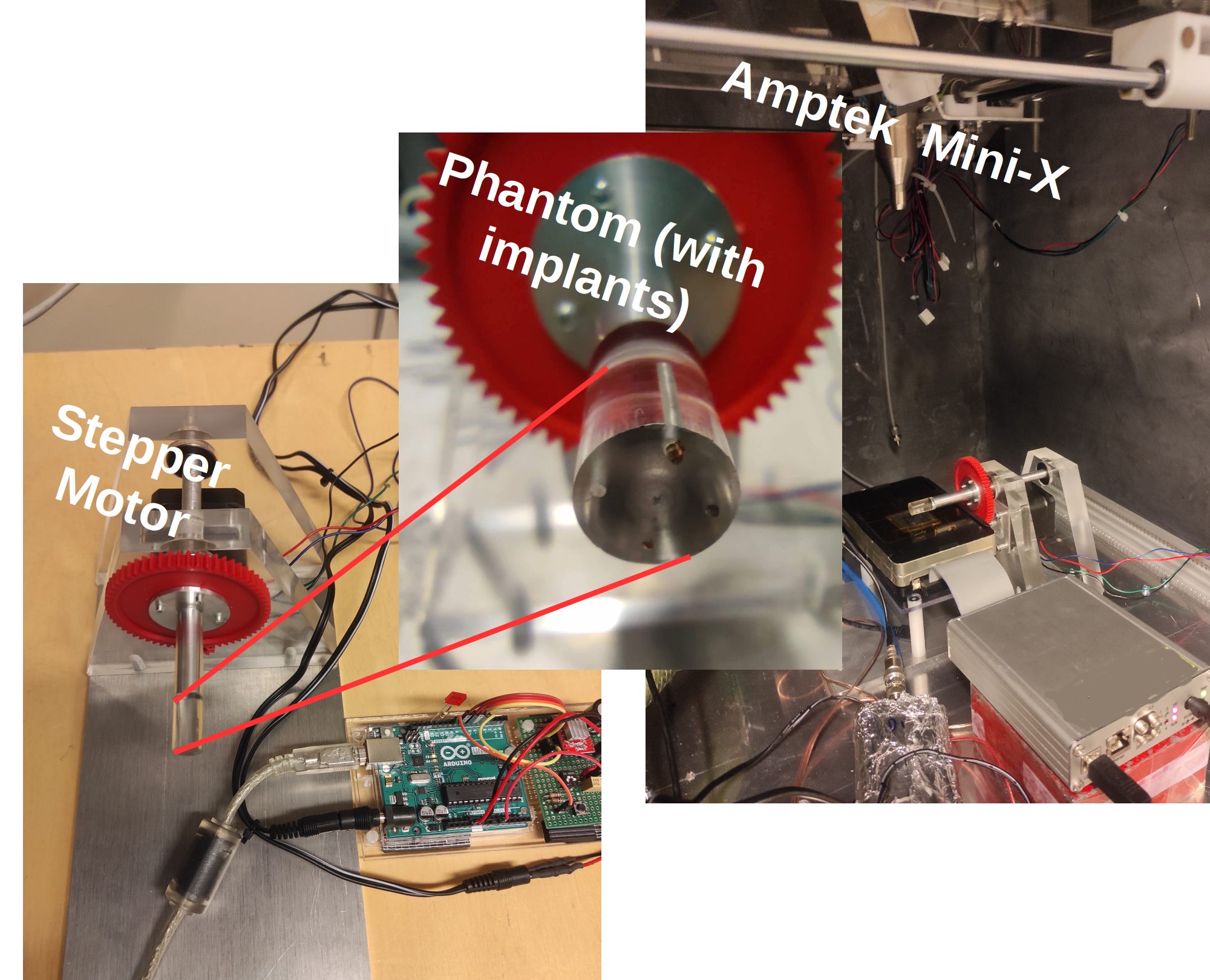}
    \caption{}
  \end{subfigure}
  \hspace{1em}
  \begin{subfigure}[b]{0.55\linewidth}
    \centering
    \includegraphics[width=\linewidth]{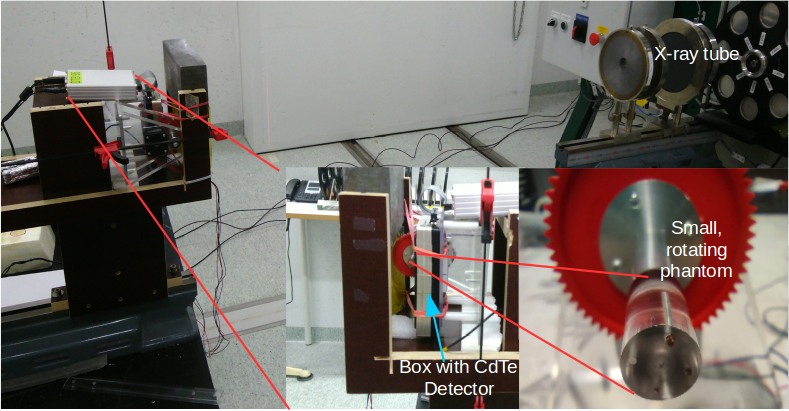}
    \caption{}
  \end{subfigure}
  \vspace*{-0.3cm}
  \caption{Measurements at HIP (a) and STUK (b). A small phantom with implants is attached to a stepper motor, controlled by an Arduino. The attenuated radiation is registered by the detector.}
  \label{MiniCT}
\end{figure}

\paragraph{Measurement setup}
At the Detector Laboratory of the Helsinki Institute of Physics (HIP) we built a small tomographic setup to perform test measurements with our detector prototypes. Fig.~\ref{MiniCT}a shows the setup, consisting of an Amptek Mini-X as X-ray source (50~kV, 80~$\mu$A), a small, rotating phantom (6\,mm diameter, acrylic glass) with three different implants (0.7\,mm; tungsten, copper, steel) and readout-electronics. The later include the PSI46digV2.1-r ASIC, flip-chip bonded to the sensors, a hybrid card and a FPGA based test board connected to a PC~\cite{Hits:2015jsa,Meier:2011zz}. The rotating is facilitated with a stepper motor (A4988 driver), controlled by an Arduino Uno. The amount of rotating steps can be defined using a customized GUI. In addition to our measurements at HIP, we performed tests with the same measurement stage at the facilities of the Radiation and Nuclear Safety Authority (STUK), where higher tube currents can be provided (cf. Fig~\ref{MiniCT}b).

At every rotation step we register the radiation attenuated by the phantom with our detector, whose individual pixels are read out in spectral mode. 
Further optimization can be achieved by clusterization of the data, summing up the charge deposition in next-neighbour pixels. 
The setup acts as a proof-of-concept for our detectors with respect to their performance as multispectral photon-counting devices. The exemplary tomographic spectrum-per-pixel data collected can be used as input to test a sophisticated reconstruction algorithm (work in progress~\cite{Purisha2019,Emzir2021}) to reconstruct the phantom as a multispectral image.

\begin{figure}[htb]
  \centering
  \begin{subfigure}[b]{0.4\linewidth}
    \centering
     \includegraphics[width=\linewidth]{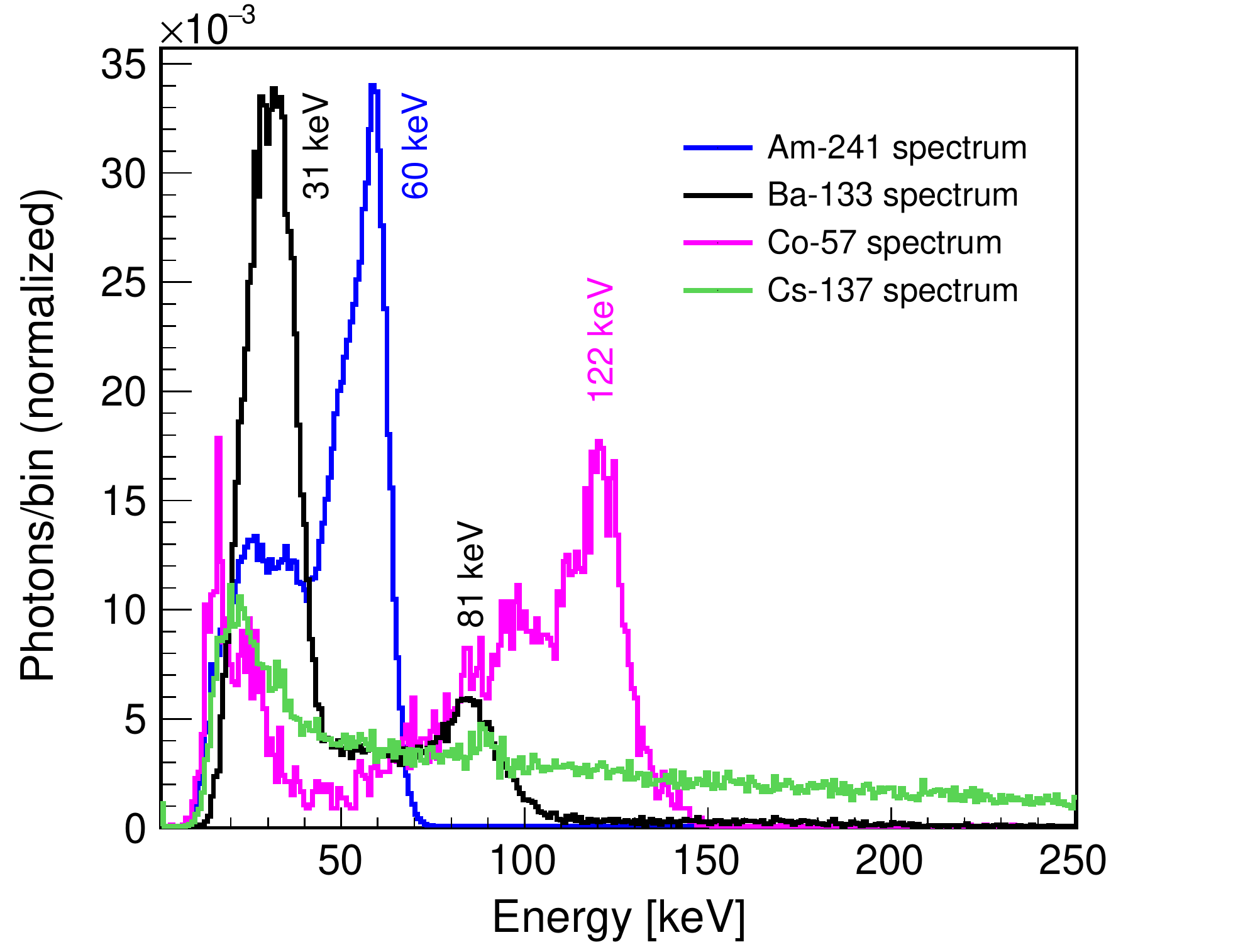}
    \caption{}
  \end{subfigure}
  \begin{subfigure}[b]{0.4\linewidth}
    \centering
    \includegraphics[width=\linewidth]{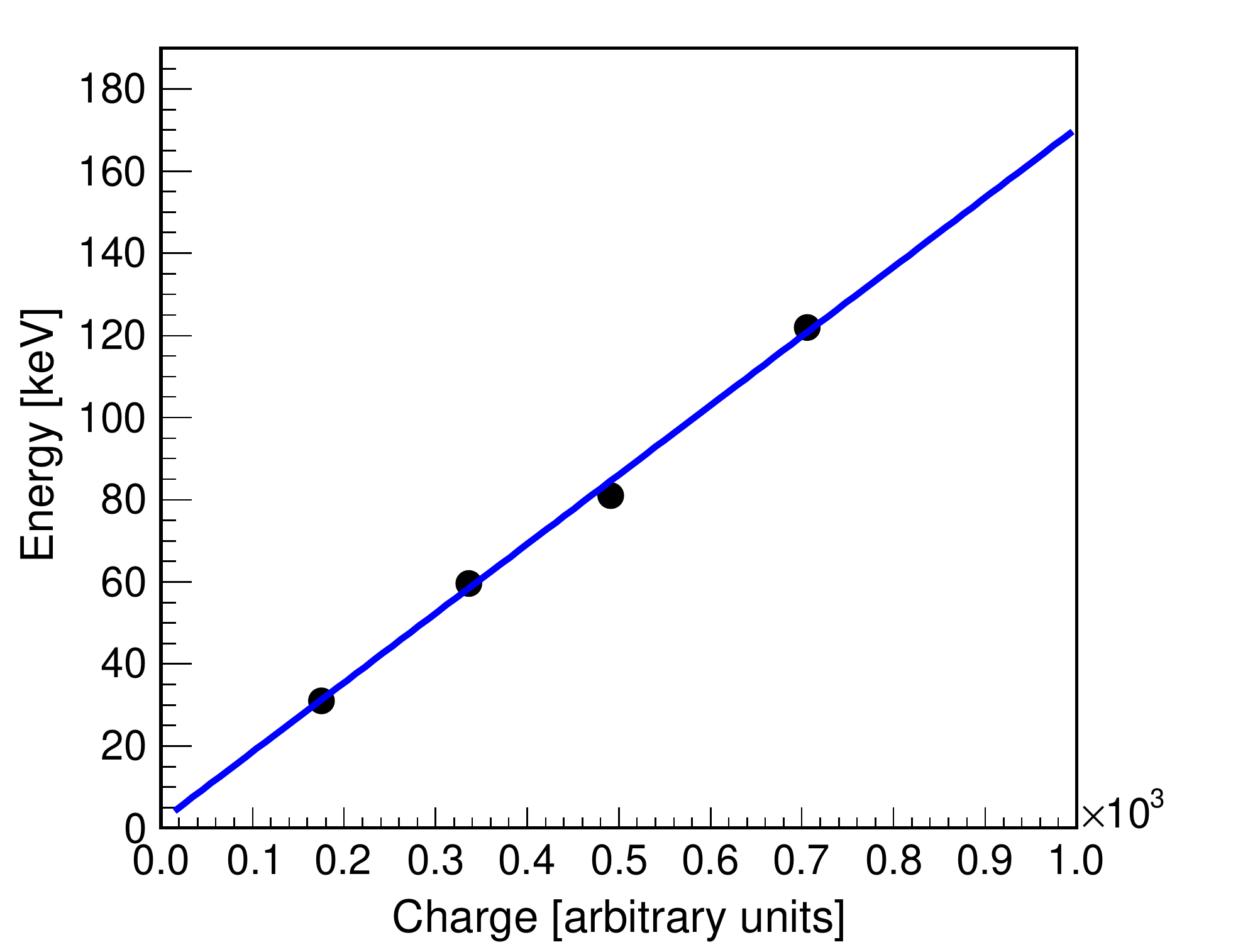}
    \caption{}
  \end{subfigure}
  \vspace*{-0.3cm}
  \caption{(a) Spectra obtained from the irradiation of a CdTe prototype detector with Am-241, Ba-133, Co-57 and Cs-137, (b) Energy calibration with sources used in (a).}
  \label{spectrum}
\end{figure}

\paragraph{Measurements with processed prototypes at HIP}
To test the response of a prototype detector to radiation from different sources, we collected spectra from Americium(Am)-241, Barium(Ba)-133, Cobalt(Co)-57 and Caesium(Cs)-137, applying a bias voltage of -300~V. At this voltage we encountered a leakage current of up to $\approx$ 0.7~\textmu A -- 1.7~\textmu A. Fig.~\ref{spectrum}a shows the measured spectra and Fig.~\ref{spectrum}b the energy calibration curve. Though the energy resolution is still sub-optimal, characteristic peaks can be seen for Am-241 (60\,keV), Ba-122 (31\,keV, 81\,keV) and Co-57 (122\,keV). Furthermore, signs of escape peaks in the region 30-40\,keV for Am-241 and 90-100\,keV for Co-57 are observed, originating from the interaction with the CdTe detector material (K$_{\alpha;\beta}$(Cd)$\approx$23;26\,keV, K$_{\alpha;\beta}$(Te)$\approx$27;21\,keV). The rather poor energy resolution at present time is partly due to electronic noise in the pre-amplification stage, because of the relatively high leakage current. In addition, we have high charge trapping in this compound semi-conductor material which leads to an effect known as hole-tailing, especially as we apply relatively low bias voltages for the 1~mm thick sensor. \\

The functionality of the rotating setup as shown in Fig.~\ref{MiniCT}a was first tested for a varying number of rotation steps. On extracting hitmaps from the respective measurement points, we can confirm the movement of the implants in accordance with the phantom during rotation (cf. next paragraph).

\paragraph{Measurements with processed prototypes at STUK}
To be able to test our prototype detectors with higher tube voltages and thus reduce the needed measurement time and enhance the possible energy spectrum, additional measurements with the rotating setup were performed at STUK. With a tube voltage of 150~kV and tube current of 0.5~mA, the leakage current during the measurement was between 1.5-7~$\mu$A.
\begin{figure}[htb]
  \centering
  \begin{subfigure}[b]{0.3\linewidth}
    \centering
     \includegraphics[width=\linewidth]{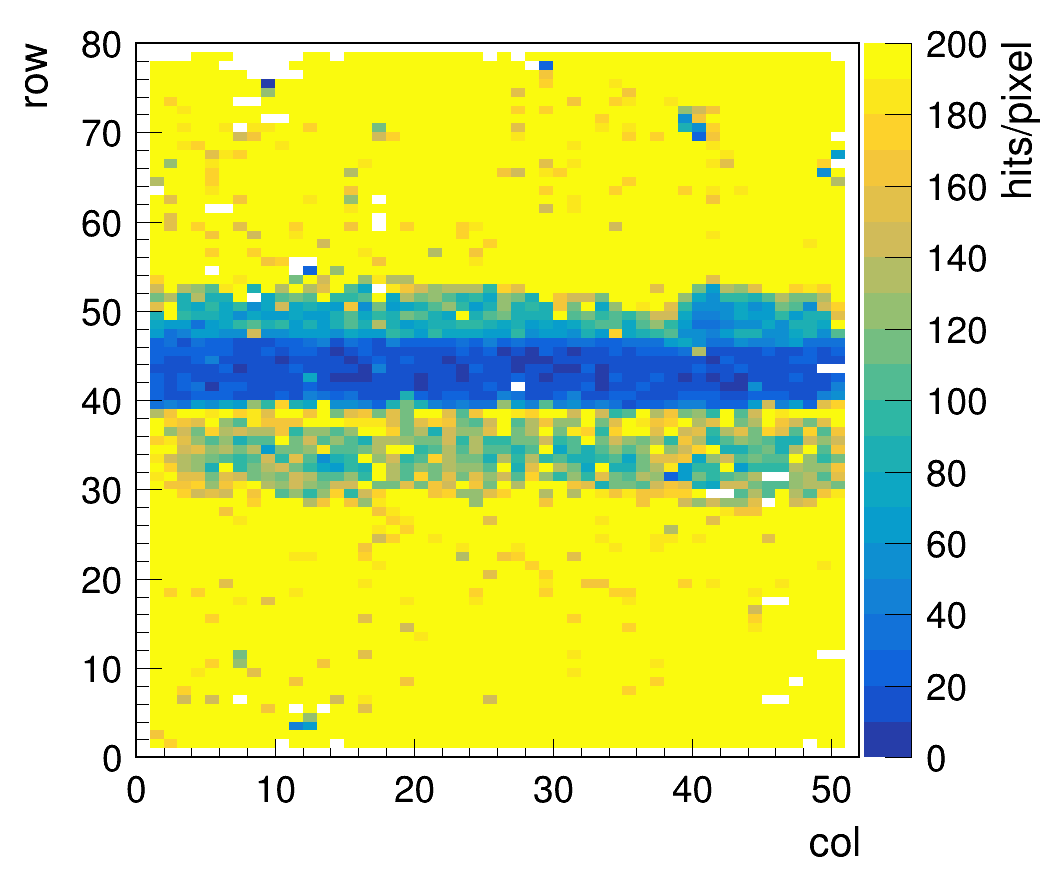}
    \caption{}
  \end{subfigure}
  \begin{subfigure}[b]{0.3\linewidth}
    \centering
    \includegraphics[width=\linewidth]{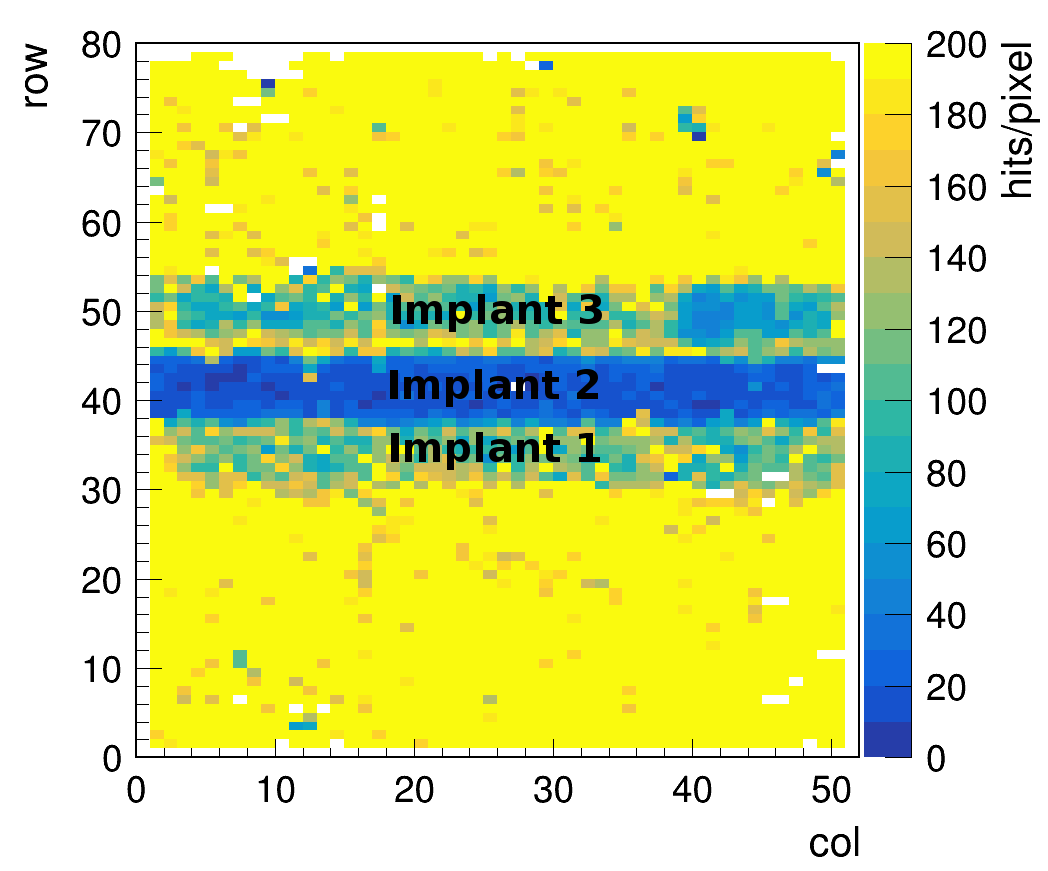}
    \caption{}
  \end{subfigure}
  \begin{subfigure}[b]{0.3\linewidth}
    \centering
    \includegraphics[width=\linewidth]{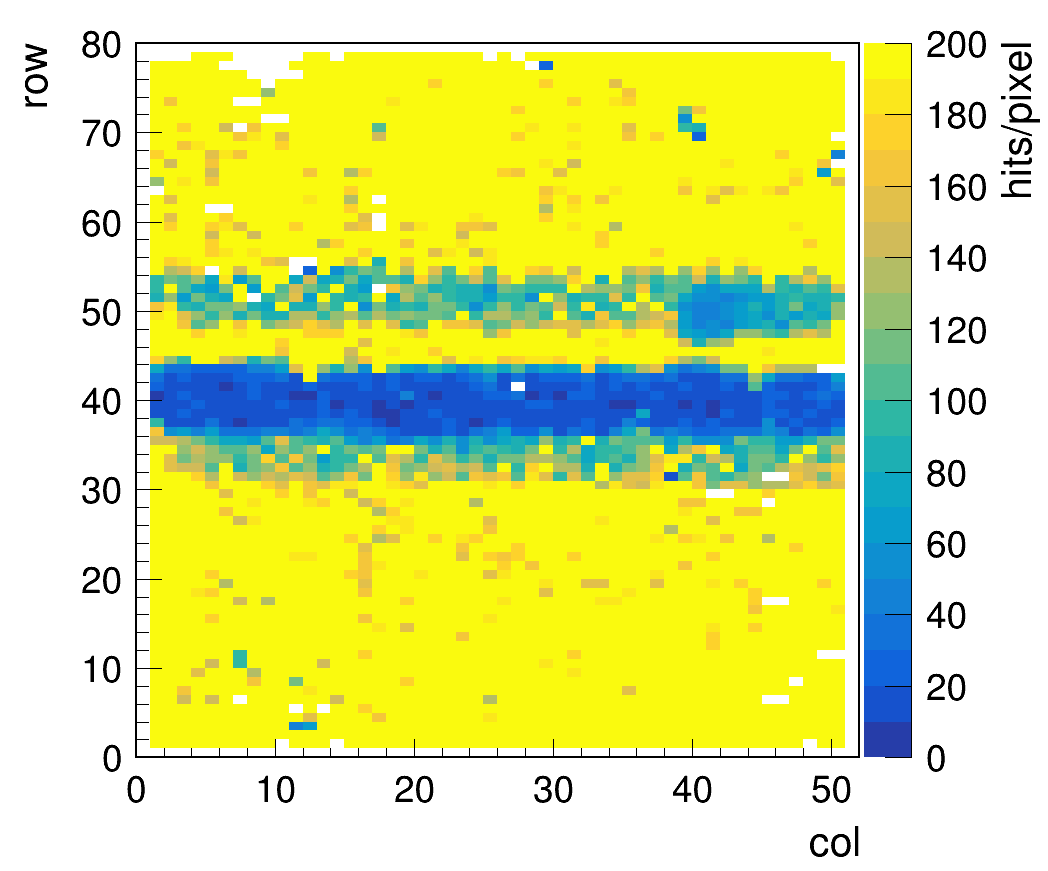}
    \caption{}
  \end{subfigure}
  \vspace*{-0.3cm}
  \caption{Hitmaps obtained from three rotation steps with a rotation angle of 9.6 degrees between subplots. \textit{Implant 1}, \textit{Implant 2} and \textit{Implant 3} mark the areas, where attenuated radiation is registered.}
  \label{hitmaps}
\end{figure}

\begin{figure}[htb]
  \centering
  \begin{subfigure}[b]{0.45\linewidth}
    \centering
    \includegraphics[width=\linewidth]{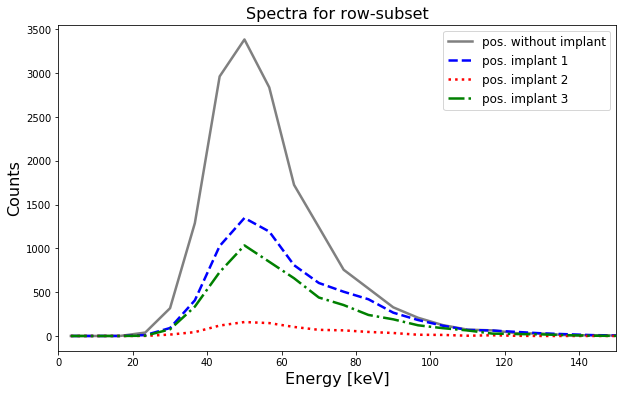}
    \caption{}
  \end{subfigure}
  \hspace{1em}
  \begin{subfigure}[b]{0.45\linewidth}
    \centering
    \includegraphics[width=\linewidth]{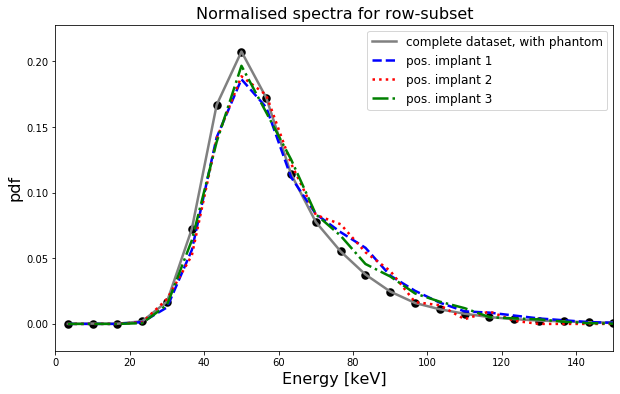}
    \caption{}
  \end{subfigure}
  \vspace*{-0.3cm}
  \caption{Comparison of spectra extracted from row-subsets of the hitmap (cf. Fig.~\ref{hitmaps}b) at rows affected by the different implants. (a) Comparing implant positions to area without implant. (b) Normalised spectra for the same row-subsets in comparison to the complete dataset (with phantom).}
  \label{spec-row-cut}
\end{figure}

The number of rotation steps was increased to 75 per rotation, corresponding to 4.8 degrees between measurements. The smaller angular difference allows for a better reconstruction of the image.
Fig.~\ref{hitmaps} shows hitmaps obtained from three measurement points during the rotation, with angular difference of 9.6 degrees. The hitmaps were obtained with the readout software pXar~\cite{pxar}, and show the amount of hits per single pixel.   Passing through an implant on the way to the detector, the radiation gets more attenuated than in the acrylic glass, leading to fewer registered hits in the respective region. As a result, we see a projection of the implants in the hitmaps, which change position in accordance with the movement of the phantom.
Fig.~\ref{spec-row-cut}a shows the spectra for row-subsets at the positions of the implants, extracted from the hitmap in Fig.~\ref{hitmaps}b. Normalising the three spectra by their area shows a slight broadening of the spectrum for higher energies (cf. Fig.~\ref{spec-row-cut}b). This is expected, as higher-energetic photons are less attenuated by the implants in these regions. Taking into account the mass attenuation coefficients for tungsten, copper and iron at about 50\,MeV~\cite{NIST} and the respective implant diameters (about 0.7\,mm), we can deduce the blue curve in Fig.~\ref{spec-row-cut}a to be related to the radiation passing the steel implant, the green curve to copper and the red curve to tungsten. The effective attenuation is a bit less than expected, as we register also counts from scattering. 

\paragraph{Reconstruction of multispectral images}
Applying standard filtered-back projection on the exemplary tomographic data described in the section before, we can as a first approach make a simple reconstruction of the phantom with visible implants. By binning our registered data into a customized number of energy bins, we can also plot and compare spectra. However, for a proper differentiation between radiation types and materials and a reconstruction of multispectral images, -- essential for the use cases of the detection system -- more advanced algorithms are necessary.\\
One possible application for a multispectral photon-counting imaging device is Boron Neutron Capture Therapy (BNCT)~\cite{KANKAANRANTA2007,Moss2014}. BNCT is an interesting alternative therapy for severe cases of cancer, but is still lacking a simultaneous monitoring and dosimetry system. In BNCT a patient is administered a drug containing Boron-10, which is primarily accumulated in cancerogeneous tissue. Next, the patient receives thermal neutron radiation targeted at the tumor location. The neutrons, which get captured by the B-10 in the cancer cells, lead to a nuclear reaction producing $\alpha$-particles which in their turn destroy the cancer cell. To keep the radiation dose applied as low as possible (but high enough as needed), the radiation dose received by the cancer cells and by the healthy cells in the vicinity must be tracked. For such a system, one would need to differentiate between the radiation types. With the multispectral CdTe detector it is in principle possible to register the prompt $\gamma$-radiation from the $^{10}$B(n,$\alpha$)$^7$Li$^{\ast}$ reaction in the cancer cell as well as the $\gamma$-radiation from the neutron background, reacting with the Cd in the detector via $^{113}$Cd(n,$\gamma$)$^{114}$Cd~\cite{Winkler2015,Brucken2020}.
For this, spectrum-per-pixel information is needed to distinguish the radiation origin (dosimetry) and furthermore reconstruct the passed material with the different tissue types (monitoring). In the framework of the MPMIB project, work is continuing to develop a sophisticated algorithm for the reconstruction of multispectral images~\cite{Purisha2019, Emzir2021}. One challenge in case of BNCT is, that no anti-scatter grid can be employed. An alternative is to simulate the scattering process and use filtered-back projection on the post-processed data.

\section{Conclusions and Outlook}
Cadmium Telluride is a promising candidate for multispectral photon-counting devices. However, the processing of CdTe crystals into full detectors is challenging. In this paper, we have presented our employed quality assessment methods that can help in sorting the CdTe material and additionally provide important insight for the consecutive processing procedure: IRM measurements are non-destructive measurements for estimating the defect density, and IV measurements can show us the preferred CdTe surface for pixelisation.
Work is continuing to compare the defect distributions to measurement results with Transient Current Technique (TCT), to study the relation between areas of higher defect densities and the charge collection efficiency of the detector~\cite{Kalliokoski2020,Golovleva2021}.\\ 
We have discussed the performance of a detector prototype in test measurements and introduced a small tomographic setup built at our laboratory, from which we extract exemplary tomographic data. This data is currently investigated closer, with the next step being the employment of a sophisticated reconstruction algorithm (under development in the MPMIB project) to reconstruct a phantom with different kinds of implants as a multispectral image. The results will give a clearer picture on how the detectors are performing with respect to multispectral imaging. A goal is building a detector array from these detectors, which can be interesting for methods such as Boron Neutron Capture Therapy (BNCT) as an online dosimetry and monitoring device.

\acknowledgments
This study was performed in the framework of the Academy of Finland project, number 314473, \textit{Multispectral photon-counting for medical imaging and beam characterization}, for which we would like to acknowledge the funding. S. Kirschenmann and S. Bharthuar thank the Magnus Ehrnrooth foundation for financial support. The measurements were performed at the joint Detector Laboratory of the University of Helsinki and the Helsinki Institute of Physics as well as at the Radiation and Nuclear Safety Authority (STUK) and we would like to thank the staff for their support.

\bibliographystyle{JHEP}
\bibliography{iworid2021_proceedings_arxiv}

\end{document}